\documentclass[fleqn,10pt]{wlscirep}
\usepackage[utf8]{inputenc}
\usepackage[T1]{fontenc}
\usepackage{amsmath}
\usepackage{color}

\DeclareMathOperator*{\argmin}{arg\,min}

\newcommand{\calU}{{\mathcal U}}

\newcommand{\bbN}{{\mathbb N}}

\newcommand{\bbR}{{\mathbb R}}

\title{Model Predictive Control for Finite Input Systems using the D-Wave Quantum Annealer}

\author[1,*]{Daisuke Inoue}
\author[1]{Hiroaki Yoshida}
\affil[1]{Toyota Central R\&D Labs., Inc., Bunkyo-ku, Tokyo, 112-0004, Japan}

\affil[*]{daisuke-inoue@mosk.tytlabs.co.jp}

\begin{abstract}
The D-Wave quantum annealer has emerged as a novel computational architecture that is attracting significant interest, but there have been only a few practical algorithms exploiting the power of quantum annealers. Here we present a model predictive control (MPC) algorithm using a quantum annealer for a system allowing a finite number of input values. Such an MPC problem is classified as a non-deterministic polynomial-time-hard combinatorial problem, and thus real-time sequential optimization is difficult to obtain with conventional computational systems. We circumvent this difficulty by converting the original MPC problem into a quadratic unconstrained binary optimization problem, which is then solved by the D-Wave quantum annealer. Two practical applications, namely stabilization of a spring-mass-damper system and dynamic audio quantization, are demonstrated. For both, the D-Wave method exhibits better performance than the classical simulated annealing method. Our results suggest new applications of quantum annealers in the direction of dynamic control problems.
\end{abstract}
\begin{document}

\flushbottom
\maketitle
%
%
\thispagestyle{empty}


\section*{Introduction}

Since \emph{quantum annealer 2000Q} was released from D-Wave Systems Inc., research on quantum computing has rapidly progressed~\cite{Kadowaki1998Quantum,Johnson2011Quantum,OGorman2015Bayesian,Venturelli2015Quantum}. 
The quantum annealer is a computational system specialized for solving combinatorial optimization problems, 
and it is expected to solve these problems with high accuracy and speed compared to classical computer systems~\cite{McGeoch2013Experimental,Denchev2016What}. 
Examples of using quantum annealers are expected to greatly expand in the future because the combinatorial optimization problem appears in various engineering fields, such as logistics \cite{sbihi2010combinatorial}, finance \cite{chinchuluun2007survey}, transportation \cite{toth2002vehicle}, and social infrastructure \cite{frangopol2007maintenance}.
Still, the class of problems that can be handled by quantum annealers is restricted compared with 
conventional computers,
and the current limitation on problem size further 
hinders the expansion of practical applications.

In this context, we shed light on the potential for application of quantum annealers to a \emph{model predictive control (MPC)}, which is one of the modern control algorithms. 
In the MPC, the future trajectory of the controlled states of a target system is predicted using a dynamic model of the target,
and the input is sequentially generated so as to minimize the evaluation function for the future state trajectory \cite{Rawlings2000Tutorial, Morari1999Model}.
Compared with other classical control methods, this method has flexibly to reflect the designer's control requirements in the evaluation function \cite{Kouro2009Model}.
The particular challenge here is that sequential optimization is computationally intensive, and thus this technique has been conventionally applied only to systems with large time constants, such as chemical plants. 
Although recent computational developments have gradually extended applicability to systems with small time constants, such as mechanical systems and electrical systems \cite{Qin2003survey,Morari1999Model,Lin2011Fast},
certain systems are still difficult to control by means of the MPC.
A system that allows only a predetermined finite number of input values (here refer to as a \emph{finite input system})
is one example of such systems.
The simplest systems consisting of on/off \cite{wonham1964optimal} input, 
and systems with finite digital bit input \cite{Bemporad1999Control} are also included in this class of systems. 
Finding the optimal input for such systems 
becomes drastically difficult as the size of the problem increases, 
because this problem is classified as a non-deterministic polynomial-time (NP)-hard combinatorial optimization problem. 
%
%
Although there are several well-known approximation methods, such as simulated annealing \cite{Suman2006survey}, tabu search \cite{Glover1998Tabu}, and genetic algorithms \cite{Mitchell1998Handbook}, the computational costs for all of these methods are still too large~\cite{vazirani2013approximation}. 

In the present study, we propose a method to solve the MPC problem by using the quantum annealing with D-Wave's quantum annealer 2000Q. 
So far, 2000Q has been used in several engineering fields \cite{OMalley2017Nonnegative,Ohzeki2018Optimization,Tran2016hybrid,Neven2008Training,Neven2009Training},
but to the best of our knowledge, 
this is first time that the 2000Q is applied to dynamic control problems in which an optimization must be obtained sequentially at high speed.
We first give a method to transform the original MPC problem into a \emph{quadratic unconstrained binary
optimization (QUBO)} problem, which is the only class of problem that the 2000Q can solve.
We then apply the proposed method to two scenarios. We first consider the \emph{stabilization of a spring-mass-damper system}, which is one of the most basic control problems to stabilize a physical system. The second scenario
 is \emph{dynamic quantization of an audio signal}, in which the audio signal is dynamically quantized considering human auditory characteristics. In both scenarios, we verify that the performance is better than the case with the simulated annealing, which is one of the classical approximation algorithms.

\section*{Results}

\subsection*{Model Predictive Control for Finite Input Systems}\label{sec:mpc}

Let us consider the following discrete-time linear system as a control target: 
\begin{align}
    \begin{split}\label{eq:dynamics}
    x(t+1) &= Ax(t) + B_1 u(t) + B_2 a(t),\\
    y(t) &= C x(t) + D u(t) + a(t),
    \end{split}
  \end{align}
where $t\in\bbN$ denotes discrete time steps; 
$x(t)\in\bbR^n$ denotes states; $u(t)\in\calU\subset \bbR^m$ denotes inputs; $a(t)\in\bbR^\ell$ denotes given signals; $y(t)\in\bbR^\ell$ denotes outputs of the system and  
$A\in\bbR^{n\times n}, B_1\in\bbR^{n\times m}, B_2\in\bbR^{n\times \ell}, C\in\bbR^{\ell\times n}$, and $D\in\bbR^{\ell \times m}$ are known constant matrices that represent the dynamics of the system.
We assume that system \eqref{eq:dynamics} is fully controllable and fully observable. 
In addition, $\calU$ is a finite set composed of $M$ vectors, and this set is defined as 
\begin{align}\label{eq:input_set}
\calU = \{u_i \mid u_i\in\bbR^m, i=1,\ldots,M\}. 
\end{align}
System \eqref{eq:dynamics} differs from the usual control system in that input $u(t)$ at each time step only takes elements of the set $\calU$.

To control system \eqref{eq:dynamics}, we design the following evaluation function:
\begin{align}\label{eq:eval_func}
H= \sum_{k=t}^{t+N-1} \left(y(k)^\top  Q y(k) + u(k)^\top  R u(k) \right),
\end{align} where $Q\in\bbR^{\ell\times\ell}$ and $R\in\bbR^{m\times m}$ are positive definite symmetric matrices called \emph{weight matrices}, which are the design parameters.
$N\in\bbN$ is also a design parameter representing the length of the evaluation section, which is called the \emph{prediction horizon}. 
The problem of finding input sequence $\mathbf{u}(t):=[u(t)\ \cdots \ u(t+N-1)]^\top$ that minimizes the evaluation function \eqref{eq:eval_func}, while observing states $x(t)$ at time $t$,  
\begin{align}\label{eq:find_inputs}
    \mathbf{u}^*(t) := \argmin_{\mathbf{u}(t)\in \calU^N} H, 
\end{align}
is called an \emph{optimal control problem}.
In the MPC, Eq.~\eqref{eq:find_inputs} is solved at each time step and the given input is applied to the control target at each time step.
This is done in the present study by using the quantum annealer 2000Q. 
We exploit the discrete nature of the input by transforming the above control problem into a \emph{quadratic unconstrained binary
optimization (QUBO)} problem.
The \emph{Methods} section provides details of the transformation. 
The evaluation function of the resulting QUBO problem takes the form 
\begin{align}\label{eq:exact_qubo_1}
  H &= \mathbf{b}(t)^\top \mathbf{J}(t) \mathbf{b}(t) + \mathbf{h}(t)^\top\mathbf{b}(t) + c'(t),
\end{align}
with the binary design variable $\mathbf{b}(t)\in \{0,1\}^{mLN}$ ($L$ is a natural number satisfying $M=2^L$). The  matrix $\mathbf{J}(t)$, vector $\mathbf{h}(t)$, and function $c'(t)$
are precisely defined in the \emph{Methods} section.
After arriving at this equivalent QUBO expression, which is compatible with quantum annealers, 
we are able to solve the optimal control problem using the quantum annealer 2000Q.

In the following discussion, we consider two scenarios to evaluate the performance of the proposed method:
\begin{itemize} 
  \item Stabilization of spring-mass-damper system: 
  For an unstable physical system, optimal stabilization control is formulated as an MPC problem. 
  \item Dynamic quantization of audio signal: 
  Audio quantization that considers human auditory characteristics is formulated as an MPC problem.
\end{itemize}
To validate the performance of the proposed method, we compare the following three solutions: 
\begin{itemize}
    \item Exact solution: A brute-force search is used to find the minimum value of Eq.~\eqref{eq:find_inputs} by enumerating all possible combinations of variables.
    \item Approximate solution using the simulated annealing: Eq.~\eqref{eq:find_inputs} is converted into a QUBO problem and solved using the simulated annealing, which is one of the classical optimization methods for combinatorial optimization. See Ref.~\citeonline{Suman2006survey} for the detailed explanation of the simulated annealing.
    \item Approximate solution using the quantum annealing: The converted QUBO problem is solved using the quantum annealer 2000Q, which is physically located in British Columbia, Canada. 
    We use HTTPS communication to command the execution of optimization from our local environment.
\end{itemize}
The \emph{Solver API} library provided by D-Wave was used for the quantum annealing method, 
and the \emph{anneal} library included in the \emph{D-Wave Ocean software package} was used 
for the simulated annealing method.
Each method was implemented using programming language Python, and executed using a Windows machine with 3.40 GHz clock frequency and 16.0 GB memory capacity.
In each method, the default parameters of the provided program were used.

\subsection*{Stabilization of Spring-Mass-Damper System}

Consider a continuous-time linear system as the control target of Eq.~\eqref{eq:dynamics}: 
\begin{align}\label{eq:example_sys}
  \dot x(\tau) &=
  \begin{bmatrix}
    0 & 1\\
    -10 & 1
  \end{bmatrix} x(\tau) +
  \begin{bmatrix}
    0\\
    1
  \end{bmatrix} u(\tau),\
  x(0) = \begin{bmatrix}
    10\\
    -10
  \end{bmatrix},\\
  y(\tau) &= x(\tau),
\end{align}
where the values are in SI units. System \eqref{eq:example_sys} is a spring-mass-damper system with a negative damper coefficient, which is a controllable but unstable system \cite{cairano2007model}.
By discretizing system \eqref{eq:example_sys} with zero-order hold of the sampling period $0.1\ \mathrm{s}$, the following discrete-time linear system is obtained:
\begin{align}
    x(t+1) &= \hat A x(t) + \hat B u(t),\\
    y(t) &= x(t),\\
    \hat A &:= \begin{bmatrix}
        0.94872313 & 0.1034271\\
        -1.03427103 & 1.05215023
    \end{bmatrix},\\
    \hat B &:= \begin{bmatrix}
        0.00512769\\
        0.1034271
    \end{bmatrix}.
\end{align}
Considering the correspondence between system \eqref{eq:example_sys} and system \eqref{eq:dynamics}, the following relation is obtained: 
\begin{align}\label{eq:relation_sys}
  A = \hat A,\ B_1 = \hat B,\ B_2 = 0,\ C = I,\ D = 0,
\end{align}
where $I$ denotes identity matrix. 
We design the weight matrices in evaluation function \eqref{eq:eval_func} as
\begin{align}
  Q = 10^5 \begin{bmatrix}
      1 & 0\\
      0 & 1
    \end{bmatrix},\
  R = 1.
\end{align}
For the set of inputs given in Eq.~\eqref{eq:input_set}, we use 
\begin{align}\label{eq:example_U1}
  \hat\calU = \{-60,-45,-30,-15,0,15,30,45\}
\end{align}
and adopt $N=6$ for the prediction horizon.

\begin{figure}[t]
  \centering
  \includegraphics[width=160mm,bb=0.000000 0.000000 825.000000 609.000000]{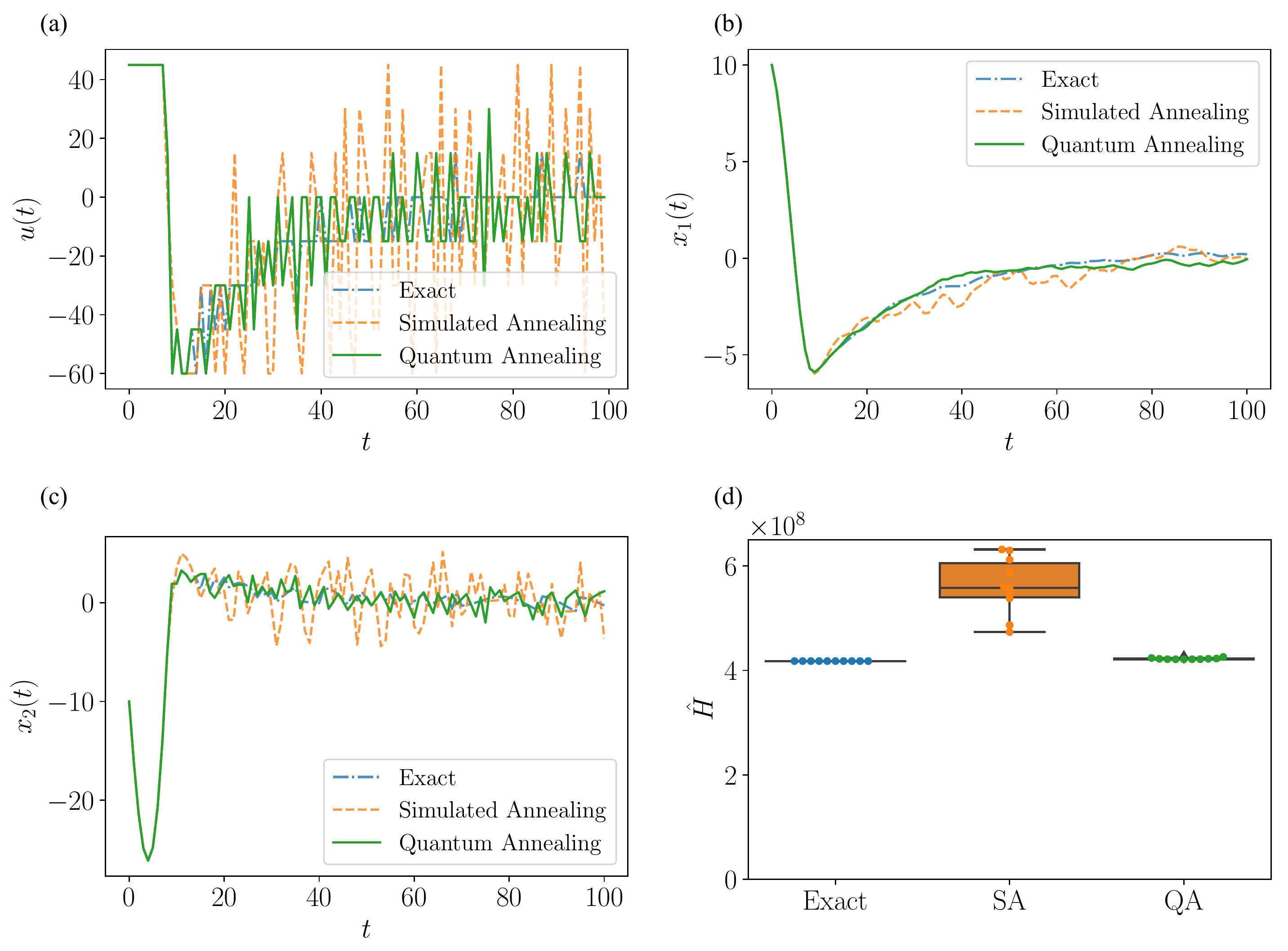}
  \caption{Result of the stabilization of a spring-mass-damper system. (a) Input $u(t)$ generated with each method. (b) Time response of the first component of the state $x(t)$. (c) Time response of the second component of the state $x(t)$. 
  In (a)--(c), the input generated with the quantum annealing is compared to the input from the simulated annealing and the brute-force search.   (d) Value of evaluation function given in Eq.~\eqref{eq:eval_func_numerical}. 
    The box-and-whisker plots are created by performing 10 numerical simulations. 
  }
  \label{fig:MPC}
\end{figure}

We show the generated input sequence in Fig.~\ref{fig:MPC}-(a) and the time response of the states in Fig.~\ref{fig:MPC}-(b) and (c).
First, for the input in Fig.~\ref{fig:MPC}-(a), the proposed method using the quantum annealing produces an input that is closer to the exact solution than that of the simulated annealing.
Also, in Fig.~\ref{fig:MPC}-(b) and (c), two states for the quantum annealing converge to the origin more rapidly than the simulated annealing.
Comparison of the simulated annealing and the quantum annealing shows that the latter method is more likely to output a solution closer to the global optimum solution.

In Fig.~\ref{fig:MPC}-(d), we show the total values of the evaluation function \eqref{eq:eval_func}, that is, 
\begin{align}\label{eq:eval_func_numerical}
  \hat H := \sum_{t=0}^{99} \left(y(t)^\top  Q y(t) + u(t)^\top  R u(t) \right).
\end{align}
%
The value of Eq.~\eqref{eq:eval_func_numerical} in the quantum annealing is smaller than the corresponding value in the simulated annealing.
The average value in each method is $4.22\times 10^{8}$ and $5.62\times 10^{8}$, respectively, and the value in the quantum annealing is suppressed to $0.75$ times that in the simulated annealing.
The standard deviations are $1.37\times 10^{8}$ and $5.19\times 10^{8}$, respectively, which means the variance of the quantum annealing solution is smaller than that of the simulated annealing.
When the quantum annealing method is compared with the exact solution method, the average value of the quantum annealing is $1.01$ times that of the exact solution.

\begin{table}[ht]
 \caption{Elapsed time for each method. }
 \label{table:time}
 \centering
  \begin{tabular}{c|c}
   \hline
    Method & Elapsed Time [s]\\
    \hline \hline
    Exact solution & $2.46\times 10^{2}$\\
    Simulated Annealing & $5.51\times 10^{-1}$\\
    Quantum Annealing (Total)& $9.68\times 10^1$\\
    QA (Computation Time)& $7.95\times 10^{-1}$\\
    QA (Annealing Time)& $2.0\times 10^{-3}$\\
   \hline
  \end{tabular}
\end{table}

Table \ref{table:time} shows the elapsed time for $100$ time steps in each method to find the input.
Quantum annealing takes $96.8\ \mathrm{s}$ to perform the optimization. 
Most of this time is occupied by communication time between the local environment and Canada, where the 2000Q is located.
The calculation time on the 2000Q is $7.95\times 10^{-1}\ \mathrm{s}$, which is $0.8\%$ of the total time, and the time required for the annealing operation is as short as $2.0\times 10^{-3}\ \mathrm{s}$.
In contrast, the calculation time for the simulated annealing and the exact solution are $5.51\times 10^{-1}\ \mathrm{s}$ and $2.46\times 10^2\ \mathrm{s}$, respectively.
To control the physical system given in Eq.~\eqref{eq:example_sys} in real time, it is desirable to generate the input at each step at a time shorter than the sampling period of $0.1\ \mathrm{s}$.
In the numerical example above, the time required for one step of input generation in each method is $2.46\ \mathrm{s}$ in the brute-force search, $5.51\times 10^{-3}\ \mathrm{s}$ in the simulated annealing, and $0.97\ \mathrm{s}$ in the quantum annealing.
Among them, only the simulated annealing achieves input generation within $0.1\ \mathrm{s}$.
However, the proposed method takes $7.95\times 10^{-3}\ \mathrm{s}$ to generate inputs when communication time is \emph{not} counted.
This suggests that if the quantum annealing can be performed in the local environment, real-time control is achieved for control targets with small time constants, such as mechanical systems and electrical systems.

\subsection*{Dynamic Quantization of Audio Signal}\label{sec:audio}

Here, we consider the problem of audio signal quantization, in which real-time input signal generation is \emph{not} as important as in physical systems.
Audio quantization is not only important from the viewpoint of data compression, but it is also indispensable when analog signals are expressed as digital signals \cite{Mitchell2004Introduction}. 
We first show that the quantization to improve the naturalness of human auditory signals is attributed to the optimal control problem for the finite input systems.

Consider the problem of quantizing the audio signal $\{a(t)\},\ t\in\bbN,\ a(t)\in\bbR$.
Let $\{u(t)\},\ t\in\bbN,\ u(t)\in\bbR$ be the signal after quantization; that is, at each time $t\in\bbN$, $u(t)\in\calU$ is satisfied for a pre-fixed set $\calU= \{u_i \mid u_i\in\bbR, i=1,\ldots,M\}$ consisting of $M$ elements.
The goal of quantizing audio signal is to reduce the data size while retaining the naturalness of the human voice content.
Human ears perceive sound with different sensitivities at different frequencies \cite{Zwicker2013Psychoacoustics}, and its auditory characteristics are expressed as \emph{perception filters} \cite{Pohlmann:2000:PDA:540747}. 
Among the available filters, we consider a linear time-invariant filter \cite{Wannamaker1992Psychoacoustically}: 
\begin{align}\label{eq:PF}
  P(z) = 1 + \bar C(zI-\bar A)^{-1}\bar B,
\end{align}
where $\bar A\in\bbR^{n\times n}, \bar B\in\bbR^{n\times 1}$, and $\bar C\in\bbR^{1\times n}$ are filter parameters. 
The quantization error after the signal passes through the perception filter is defined as
\begin{align}\label{eq:audio_error}
  y(t) := P(z) (a(t)-u(t)),
\end{align}
and the evaluation function is defined as
\begin{align}\label{eq:eval_PF}
  H &= \sum_{k=t}^{t+N-1} (y(k))^2.
\end{align}

Given perception filter $P$ and audio signals $\{a(t)\}$, the problem of finding an input that minimizes the evaluation function is represented as an optimal control problem.
Using Z-transform, Eq.~\eqref{eq:audio_error} is expressed as the following state-space model: 
\begin{align}
  \begin{split}\label{eq:PF_ss}
  x(t+1) &= \bar A x(t) + \bar B(a(t)-u(t)),\\
  y(t) &= \bar C x(t) + a(t)-u(t).
  \end{split}
\end{align}
The correspondence between system \eqref{eq:PF_ss} and system \eqref{eq:dynamics} 
is as follows: 
\begin{align}\label{eq:relation_sys_PF}
  A = \bar A,\ B_1 = -\bar B,\ B_2 = \bar B,\ C = \bar C, D = -1, 
\end{align}
and the correspondence between evaluation function \eqref{eq:eval_PF} and \eqref{eq:eval_func} is as follows:
\begin{align}\label{eq:relation_eval}
  Q = 1,\ R = 0.
\end{align}
In this way, the problem of quantizing the signal to minimize evaluation function \eqref{eq:eval_PF} is expressed as an optimal control problem.

%
\begin{figure}[t]
  \centering
  \includegraphics[width=160mm,bb=0.000000 0.000000 825.000000 314.000000]{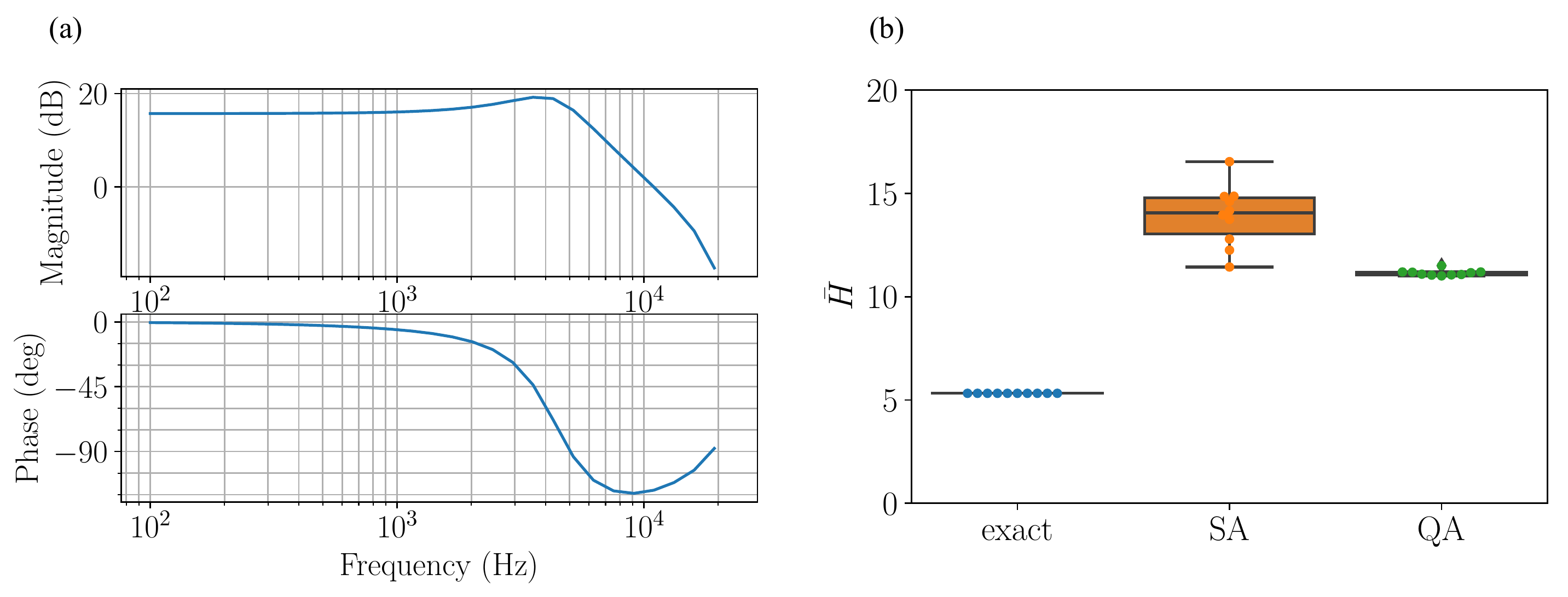}
  \caption{Result of dynamic audio quantization. (a) Bode plot of perception filter given in Eq.~\eqref{eq:PF_1}. 
  (b) Value of evaluation function given in Eq.~\eqref{eq:eval_PF_numerical}. 
  The solution from the quantum annealing (green) is compared to that from the simulated annealing (orange) and exact solution (blue).
  The box-and-whisker plot is created by performing 10 numerical simulations. }
  \label{fig:MPC_audio}
\end{figure}

For the numerical simulation, we use the following filter 
proposed in Ref.~\citeonline{Wannamaker1992Psychoacoustically}
(with a sampling frequency of $44.1$ kHz):
\begin{align}\label{eq:PF_1}
  P(z) = 1 + z^{-1}\frac{2.245 - 0.664 z^{-1}}{1-1.335z^{-1}+0.644z^{-2}}.
\end{align}
The Bode plot for this filter is depicted in Fig.~\ref{fig:MPC_audio}-(a).
Eq.~\eqref{eq:PF_1} is essentially a low-pass filter, which also enhances frequencies in the band around $4$ kHz. The parameters when expressing Eq.~\eqref{eq:PF_1} in the state space are as follows:
\begin{align}
  \bar A &= \begin{bmatrix}
  1.335 &  -0.644\\
  1.0   &  0
  \end{bmatrix},\\
  \bar B &= \begin{bmatrix}
  1.0\\
  0
  \end{bmatrix},\\
  \bar C &= \begin{bmatrix}
  2.245 & -0.664
  \end{bmatrix}.
\end{align}
We adopt the following input sets:
\begin{align}\label{eq:example_U2}
  \bar\calU = \{-1.0, -0.5, 0, 0.5 \}.
\end{align}
We use $N=10$ for the prediction horizon.
For the pre-quantized audio signal $\{a(t)\}$, we use a random number sequence of 100 steps following a uniform distribution on $[-1,1)$. 

Comparison of the values obtained by evaluating Eq.~\eqref{eq:eval_PF} over the duration with three methods, that is,
\begin{align}\label{eq:eval_PF_numerical}
  \bar H &:= \sum_{t=0}^{99} (y(t))^2,
\end{align}
is shown in Fig.~\ref{fig:MPC_audio}-(b).
Comparison of simulated and quantum annealing results shows that the value of Eq.~\eqref{eq:eval_PF_numerical} is smaller for the quantum annealing.
The average values are $13.9$ and $11.2$ respectively, and the value of quantum annealing is $0.81$ times that of the simulated annealing.
The quantum annealing solutions have less variance; the variances of the simulated and quantum annealing are $1.94$ and $1.78\times 10^{-2}$, respectively.
When comparing the exact solution and the quantum annealing, the average value from the quantum annealing is as $2.1$ times that of the exact solution, which is a conservative result when compared with the previous example. 

\begin{table}[ht]
  \caption{Elapsed time for each method. }
  \label{table:time-audio}
  \centering
   \begin{tabular}{c|c}
    \hline
     Method & Elapsed Time [s]\\
     \hline \hline
     Exact solution & $8.61\times 10^2$\\
     Simulated Annealing & $3.74\times 10^{-1}$\\
     Quantum Annealing (Total)& $8.6\times 10^1$\\
     QA (Computation Time)& $7.91\times 10^{-1}$\\
     QA (Annealing Time)& $2.0\times 10^{-3}$\\
    \hline
   \end{tabular}
 \end{table}

Table \ref{table:time-audio} shows the elapsed time required for the optimization after 100 steps in each method.
The quantum annealing takes $8.6\times 10^1\ \mathrm{s}$ to perform the optimization. 
As mentioned in the previous example, HTTPS communication accounts for most of the time, and the 2000Q calculation time is $7.91\times 10^{-1}\ \mathrm{s}$, only $0.9$\% of the total time.
In addition, the time taken for the annealing itself is even shorter, $2.0\times 10^{-3}\ \mathrm{s}$.
In contrast, the calculation time for the simulated annealing and the exact solution is $3.74\times 10^{-1} \mathrm{s}$ and $8.61\times 10^2\ \mathrm{s}$, respectively.
The proposed method using the quantum annealing enables higher-speed quantization than the brute-force search and higher performance than the simulated annealing.

\section*{Discussion}

\begin{figure}[t]
  \centering
  \includegraphics[width=85mm,bb=0.000000 0.000000 428.323765 275.387403]{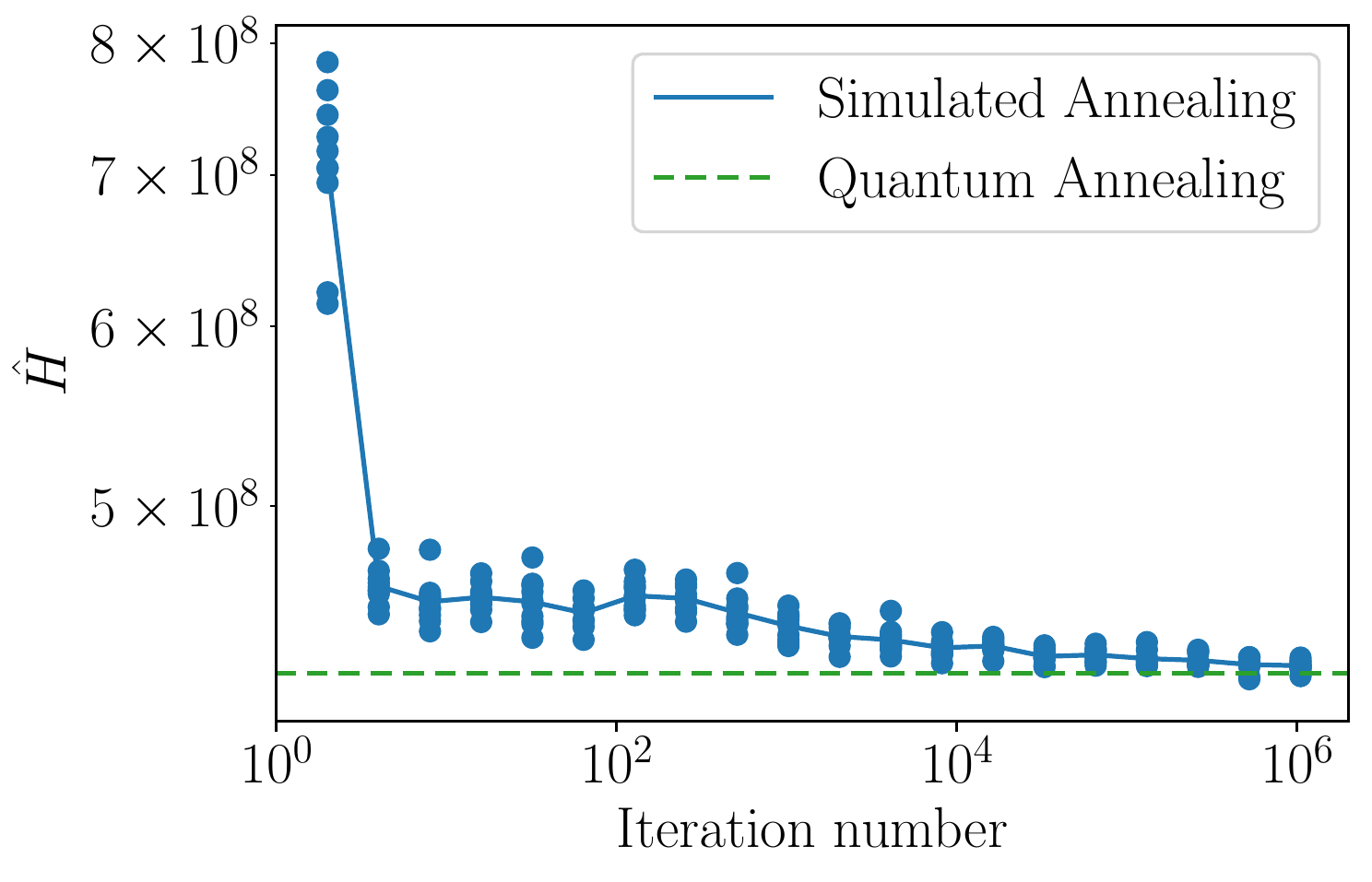}
  \caption{
  Iteration number for the simulated annealing versus the value of the evaluation function
  given in Eq.~\eqref{eq:eval_func_numerical}.
  At each value of iteration number, 10 computations are performed (blue dot) and the average value is plotted (blue solid line).
  The corresponding value of the evaluation function obtained with the quantum annealing method is also shown by the green dashed line.
  }
  \label{fig:iter_num}
\end{figure}
{
In the 2000Q, the evaluation function is minimized using a natural phenomenon called a quantum effect, and the solution is \emph{not} updated by the algorithmic iteration,
and indeed this is the essential difference from the classical simulated annealing method.
The latter employs an iterative algorithm, 
where the neighborhoods of the current solution are examined at each time step,
with probabilistically judging whether to update the solution or to remain in the current state.
The number of iterations is a design parameter (in the numerical experiment above, the library default value of $1000$ is used as the number of iterations).
Clearly the larger the number of iterations is, the more accurate the obtained solution is,
but the longer the required computational time becomes.
To examine this trade-off, we show in Fig.~\ref{fig:iter_num} the relationship between the iteration number of the simulated annealing method and the quality of the solution in the spring-mass-damper system.
The value of the evaluation function decreases as the iteration number is increased, and 
about one million steps are required to reach the level of the quantum annealing result.
The average computational time of $ 1.0 \times 10^6 $ iterations was $ 3.9 $ seconds,
which is $4.9$ times longer than 
$7.95\times 10^{-1}$ second, the time required for the quantum annealing.}

{ 
We also mention the restriction of the current quantum annealing using the 2000Q,
which is equipped no more than 2,048 qubits.
Each qubit is \emph{not} coupled with all the other qubits, but instead the assemblage has a \emph{chimera structure}, in which closely connected 8-bit units are arranged vertically and horizontally \cite{Boothby:2016:FCM:2877060.2877142}. 
For this reason, the variables of a given QUBO problem cannot be directly assigned to physical qubits. 
The method of converting the given graph structure to the chimera structure is called \emph{minor embedding}, which is realized by expressing one logical variable with strongly coupled multiple physical qubits. 
The embedding problem itself is actually an NP-hard problem, and we used the API that D-Wave provides to perform embedding with heuristic algorithms \cite{Cai2014practical}.
Because multiple physical bits are sacrificed in minor embedding, the effective number of variables decreases depending on the \emph{density} of the original problem, that is, the number of non-zero elements of a QUBO matrix.
In the chimera structure, $N^2/4$ physical qubits are needed for representing a fully connected $N$-variable problem, which means that the maximum number of variables that the 2000Q can use is as small as 64 when the original problem has a fully connected structure. This implies that $mLN\le 64$ must be satisfied for the dimension $\mathbf{b}(t)$ in Eq.~\eqref{eq:exact_qubo_1}.}
\begin{figure}[t]
  \centering
  \includegraphics[width=160mm,bb=0.000000 0.000000 825.000000 314.000000]{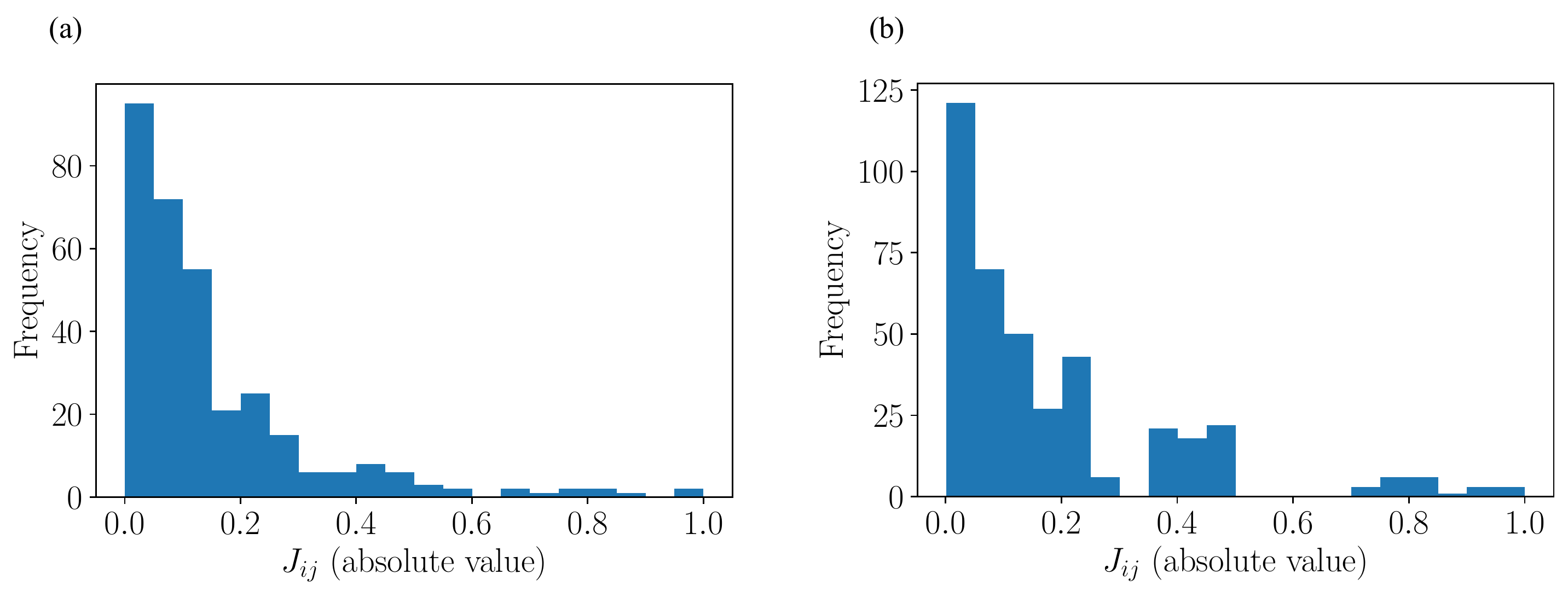}
  \caption{Histogram of the absolute value of the QUBO matrix. (a) Stabilization of spring-mass-damper system. (b) Dynamic audio quantization. }
  \label{fig:hist}
\end{figure}
In addition to the limited number of variables, it is known that the quality of the solution degrades when the original problem has a dense structure\cite{Hamerlyeaau0823}.
To check the sparseness of our formulated problem, we examine the value of all components of the QUBO matrix [$J_{ij}$ in Eq.~\eqref{eq:QUBO_matrix}] and plot them as histograms in 
Fig.~\ref{fig:hist}.
We show the result of stabilization of the spring-mass-damper system in Fig.~\ref{fig:hist}-(a) and of dynamic audio quantization in Fig.~\ref{fig:hist}-(b). 
In both cases, we plot the absolute value of the QUBO matrix and normalize them such that $\max_{i,j} J_{ij}=1$.
Many components have values close to zero, and
qualitatively, $51$~\% of components are smaller than $0.1$ for both cases, and the distribution is nearly exponential.
This implies that the problem considered in this study is relatively sparse and matches the structure of the problems that the 2000Q can solve.
Still, the sparse chimera structure limits the number of variables and the accuracy of the solution. 
D-Wave Systems Inc. recognizes this and has stated that they will adopt a new graph structure that increases the couplers per qubit in the next generation of machines \cite{W.Johnson2018Future}. 
Also, they plan to double the number of qubits in the next 2 years.
Hence, the restriction on the problem size will soon be relaxed and the accuracy of solution will be improved in the future.
We expect D-Wave's next-generation machine to be capable of solving MPC problems with higher performance than the current generation machines. 

One of the important results obtained in the present study 
is short effective computational time of the quantum annealing compared with that of the simulated annealing, 
as shown in Tables~\ref{table:time} and \ref{table:time-audio}.
In the MPC, optimization must be performed sequentially at each time step for input calculation,
and the performance deteriorates significantly when the input generation is \emph{not} synchronized with the control cycle\cite{zhang2001stability}.
If the problem at each time step is NP-hard, as in this study, this performance degradation becomes a critical problem.
Our results of using the quantum annealing to obtain a solution within a control period suggest the possibility of expanding the classes of systems that can be controlled in real time.

\section*{Methods}

\subsection*{Transformation of MPC to QUBO}
We first represent evaluation function \eqref{eq:eval_func} as a quadratic form for the input vector $\mathbf{u}(t)$.
First, the dynamics of the system \eqref{eq:dynamics} is represented as
\begin{align}\label{eq:large_dynamics}
  \mathbf{y}(t) = \mathbf{A} x(t) + \mathbf{B}_1 \mathbf{u}(t) + \mathbf{B}_2 \mathbf{a}(t),
\end{align}
with the following definitions:
\begin{align}
  \mathbf{y}(t) &:= \begin{bmatrix}
    y(t)\\
    \vdots\\
    y(t+N-1)
  \end{bmatrix}\in\bbR^{\ell N},\\
  \mathbf{a}(t) &:=\begin{bmatrix}
    a(t)\\
    \vdots\\
    a(t+N-1)
  \end{bmatrix}\in\bbR^{\ell N},\\
  \mathbf{A} &:=\begin{bmatrix}
    C\\
    CA\\
    \vdots\\
    CA^{N-1}
  \end{bmatrix}\in\bbR^{\ell N\times n},\\
  \mathbf{B}_1 &:=\begin{bmatrix}
    D & 0 & \cdots & 0\\
    C B_1 & D & \ddots & \vdots\\
    \vdots & \ddots & \ddots & 0\\
    C A^{N-2} B_1 & \cdots & C B_1 & D
  \end{bmatrix}\in\bbR^{\ell N\times mN},\\
  \mathbf{B}_2 &:=\begin{bmatrix}
    1 & 0 & \cdots & 0\\
    C B_2 & 1 & \ddots & \vdots\\
    \vdots & \ddots & \ddots & 0\\
    C A^{N-2} B_2 & \cdots & C  B_2 & 1
  \end{bmatrix}\in\bbR^{N\times N}.
\end{align}
Also, evaluation function \eqref{eq:eval_func} is expressed as
\begin{align}\label{eq:large_eval_func}
  H = \mathbf{x}(t)^\top  \mathbf{Q} \mathbf{x}(t) + \mathbf{u}(t)^\top \mathbf{R}\mathbf{u}(t)
\end{align}
by defining
\begin{align}
  \mathbf{Q} &:= \begin{bmatrix}
    Q & 0 & \cdots & 0\\
    0 & Q & \ddots & \vdots\\
    \vdots & \ddots & \ddots & 0\\
    0 & 0 & \cdots & Q
  \end{bmatrix}\in\bbR^{\ell N\times \ell N},\\
  \mathbf{R} &:= \begin{bmatrix}
    R & 0 & \cdots & 0\\
    0 & R & \ddots & \vdots\\
    \vdots & \ddots & \ddots & 0\\
    0 & 0 & \cdots & R
  \end{bmatrix}\in\bbR^{mN\times mN}.
\end{align}
By substituting Eq.~\eqref{eq:large_dynamics} into Eq.~\eqref{eq:large_eval_func}, we obtain the following quadratic form: 
\begin{align}\label{eq:eval_quad}
  \begin{split}
      H &= \mathbf{u}(t)^\top (\mathbf{B}_1^\top \mathbf{Q}\mathbf{B}_1 + \mathbf{R})\mathbf{u}(t) + 2( \mathbf{A} x(t) + \mathbf{B}_2\mathbf{a}(t)  )^\top\mathbf{Q}\mathbf{B}_1 \mathbf{u}(t)  + c(t),
  \end{split}
\end{align}
where we assumed that $Q$ is symmetrical and that $c(t)$ is a constant independent of $\mathbf{u}(t)$.

The need to search the set $\calU^N$ makes it difficult to find the minima $\mathbf{u}(t)$ of the right side of Eq.~\eqref{eq:eval_quad}. 
$\calU^N$ consists of $M^N$ elements (where $M$ is the number of elements of the set $\calU$ and $N$ is the control horizon), and as $N$ and $M$ increases, the search becomes impractical.
For this reason, when $M$ and $N$ are large, it is necessary to give up an exact solution and instead adopt an approximate solution.
This study explore the quantum annealer as one of the approximation methods.
However, because this method only accepts QUBO problems, we next need to convert Eq.~\eqref{eq:eval_quad} to a QUBO problem. 
For the set $\calU$ defined by Eq.~\eqref{eq:input_set}, we add the following assumptions: 
let $M$ (the number of elements of sets $\calU$) satisfy $M=2^L$ for some natural number $L\in\bbN$, and for all $i$, let $u_i\in\calU$ satisfy 
\begin{align}\label{eq:binary_encoding}
  u_i = \begin{bmatrix}
    K_1 \left(-2^{L-1} b_{1,L} + \sum_{\ell=1}^{L-1} 2^{\ell-1} b_{1,\ell}\right)\\
    \vdots\\
    K_m \left(-2^{L-1} b_{m,L} + \sum_{\ell=1}^{L-1} 2^{\ell-1} b_{m,\ell}\right)\\
  \end{bmatrix}, 
\end{align}
where $b_{j,\ell}\in\{0,1\}\ (j=1,\ldots,m,\ \ell=1,\ldots,L)$ is a binary variable that takes a value of either $0$ or $1$ and $K_j\in\bbR\ (j=1,\ldots,m)$ is a scaling parameter. 
For example, the set $\hat\calU$ in Eq.~\eqref{eq:example_U1} and $\bar\calU$ in Eq.~\eqref{eq:example_U2} are expressed by Eq.~\eqref{eq:binary_encoding} using $m=1, L=3$, and $K_1=15$ and $m=1, L=2$, and $K_1=0.5$, respectively. 
{
The assumption that the input set $ \calU $ is expressed as Eq.~\eqref{eq:binary_encoding}
is made to express the evaluation function \eqref{eq:eval_func} in a QUBO form.
We note that the expression in terms of Eq.~\eqref{eq:binary_encoding} 
restricts the input set to the discrete values with a regular interval.
Nevertheless, this is valid for various practical systems,
and wider classes of problems should be dealt with by means of appropriate variable transformations, which is indeed included in our future studies.}

Eq.~\eqref{eq:binary_encoding} can be viewed as a linear mapping from $\{0,1\}^{mL}$ to $\calU$, and its expression as a matrix is 
\begin{align}
  u_i &= E b,\label{eq:binary_to_discrete}\\
  E &:= \begin{bmatrix}
    E_1 & 0 & \cdots & 0\\
    0 & E_2 & \ddots & \vdots\\
    \vdots & \ddots & \ddots & 0\\
    0 & 0 & \cdots & E_m
  \end{bmatrix}\in\bbR^{m\times mL},\\
  E_i &:= \begin{bmatrix}
    K_i & \cdots & K_i 2^{L-1} & -K_i 2^{L}
  \end{bmatrix}\in\bbR^{1\times L},\\
  b &:= \begin{bmatrix}
    b_{1,1} & \cdots & b_{1,L} & \cdots & b_{m,1} & \cdots & b_{m ,L}
  \end{bmatrix}^\top\nonumber\in\{0,1\}^{mL}. 
\end{align}
In addition, by defining following block matrices,
\begin{align}
  \mathbf{E} &:= \begin{bmatrix}
    E & 0 & \cdots & 0\\
    0 & E & \ddots & \vdots\\
    \vdots & \ddots & \ddots & 0\\
    0 & 0 & \cdots & E
  \end{bmatrix}\in\bbR^{mN\times mLN},\\
  \mathbf{b}(t) &:=\begin{bmatrix}
    b(t)\\
    \vdots\\
    b(t+N-1)
  \end{bmatrix}\in\{0,1\}^{mLN}\label{eq:expand_b}, 
\end{align}
Eq.~\eqref{eq:eval_quad} can be written as a QUBO as follows:
\begin{align}\label{eq:exact_qubo}
  H &= \mathbf{b}(t)^\top \mathbf{J}(t) \mathbf{b}(t) + \mathbf{h}(t)^\top\mathbf{b}(t) + c'(t),
\end{align}
where $c'(t)$ is a constant independent of $\mathbf{b}(t)$ and $\mathbf{J}(t)$ and $\mathbf{h}(t)$ are defined as 
\begin{align}\label{eq:QUBO_matrix}
    \mathbf{J}(t) &:= \mathbf{E}^\top (\mathbf{B}_1^\top \mathbf{Q}\mathbf{B}_1 + \mathbf{R})\mathbf{E},\\
    \mathbf{h}(t) &:= 2( \mathbf{A} x(t) + \mathbf{B}_2\mathbf{a}(t)  )^\top\mathbf{Q}\mathbf{B}_1 \mathbf{E}.
\end{align}
In this way, the problem of determining the right-hand side of Eq.~\eqref{eq:eval_func} is expressed as a QUBO that minimizes Eq.~\eqref{eq:exact_qubo} on $\mathbf{b}(t)$.
Once the QUBO is solved by the 2000Q, the original input $\mathbf{u}(t)$ is recovered by substituting solution $\mathbf{b}(t)$ into Eq.~\eqref{eq:binary_to_discrete}.

\bibliography{main}

\begin{thebibliography}{10}
\urlstyle{rm}
\expandafter\ifx\csname url\endcsname\relax
  \def\url#1{\texttt{#1}}\fi
\expandafter\ifx\csname urlprefix\endcsname\relax\def\urlprefix{URL }\fi
\expandafter\ifx\csname doiprefix\endcsname\relax\def\doiprefix{DOI: }\fi
\providecommand{\bibinfo}[2]{#2}
\providecommand{\eprint}[2][]{\url{#2}}

\bibitem{Kadowaki1998Quantum}
\bibinfo{author}{Kadowaki, T.} \& \bibinfo{author}{Nishimori, H.}
\newblock \bibinfo{journal}{\bibinfo{title}{Quantum {{Annealing}} in the
  {{Transverse Ising Model}}}}.
\newblock {\emph{\JournalTitle{Physical Review E}}}
  \textbf{\bibinfo{volume}{58}}, \bibinfo{pages}{5355} (\bibinfo{year}{1998}).

\bibitem{Johnson2011Quantum}
\bibinfo{author}{Johnson, M.~W.} \emph{et~al.}
\newblock \bibinfo{journal}{\bibinfo{title}{Quantum {{Annealing}} with
  {{Manufactured Spins}}}}.
\newblock {\emph{\JournalTitle{Nature}}} \textbf{\bibinfo{volume}{473}},
  \bibinfo{pages}{194} (\bibinfo{year}{2011}).

\bibitem{OGorman2015Bayesian}
\bibinfo{author}{O'Gorman, B.}, \bibinfo{author}{Babbush, R.},
  \bibinfo{author}{{Perdomo-Ortiz}, A.}, \bibinfo{author}{{Aspuru-Guzik}, A.}
  \& \bibinfo{author}{Smelyanskiy, V.}
\newblock \bibinfo{journal}{\bibinfo{title}{Bayesian network structure learning
  using quantum annealing}}.
\newblock {\emph{\JournalTitle{The European Physical Journal Special Topics}}}
  \textbf{\bibinfo{volume}{224}}, \bibinfo{pages}{163--188}
  (\bibinfo{year}{2015}).

\bibitem{Venturelli2015Quantum}
\bibinfo{author}{Venturelli, D.}, \bibinfo{author}{Marchand, D.~J.} \&
  \bibinfo{author}{Rojo, G.}
\newblock \bibinfo{journal}{\bibinfo{title}{Quantum annealing implementation of
  job-shop scheduling}}.
\newblock {\emph{\JournalTitle{arXiv preprint arXiv:1506.08479}}}
  (\bibinfo{year}{2015}).

\bibitem{McGeoch2013Experimental}
\bibinfo{author}{McGeoch, C.~C.} \& \bibinfo{author}{Wang, C.}
\newblock \bibinfo{title}{Experimental {{Evaluation}} of an {{Adiabiatic
  Quantum System}} for {{Combinatorial Optimization}}}.
\newblock In \emph{\bibinfo{booktitle}{Proceedings of the {{ACM International
  Conference}} on {{Computing Frontiers}}}}, \bibinfo{pages}{23}
  (\bibinfo{organization}{ACM}, \bibinfo{year}{2013}).

\bibitem{Denchev2016What}
\bibinfo{author}{Denchev, V.~S.} \emph{et~al.}
\newblock \bibinfo{journal}{\bibinfo{title}{What {{Is}} the {{Computational
  Value}} of {{Finite}}-{{Range Tunneling}}?}}
\newblock {\emph{\JournalTitle{Physical Review X}}}
  \textbf{\bibinfo{volume}{6}}, \bibinfo{pages}{031015} (\bibinfo{year}{2016}).

\bibitem{sbihi2010combinatorial}
\bibinfo{author}{Sbihi, A.} \& \bibinfo{author}{Eglese, R.~W.}
\newblock \bibinfo{journal}{\bibinfo{title}{Combinatorial optimization and
  green logistics}}.
\newblock {\emph{\JournalTitle{Annals of Operations Research}}}
  \textbf{\bibinfo{volume}{175}}, \bibinfo{pages}{159--175}
  (\bibinfo{year}{2010}).

\bibitem{chinchuluun2007survey}
\bibinfo{author}{Chinchuluun, A.} \& \bibinfo{author}{Pardalos, P.~M.}
\newblock \bibinfo{journal}{\bibinfo{title}{A survey of recent developments in
  multiobjective optimization}}.
\newblock {\emph{\JournalTitle{Annals of Operations Research}}}
  \textbf{\bibinfo{volume}{154}}, \bibinfo{pages}{29--50}
  (\bibinfo{year}{2007}).

\bibitem{toth2002vehicle}
\bibinfo{author}{Toth, P.} \& \bibinfo{author}{Vigo, D.}
\newblock \emph{\bibinfo{title}{The Vehicle Routing Problem}}
  (\bibinfo{publisher}{{SIAM}}, \bibinfo{year}{2002}).

\bibitem{frangopol2007maintenance}
\bibinfo{author}{Frangopol, D.~M.} \& \bibinfo{author}{Liu, M.}
\newblock \bibinfo{journal}{\bibinfo{title}{Maintenance and management of civil
  infrastructure based on condition, safety, optimization, and life-cycle
  cost{${_\ast}$}}}.
\newblock {\emph{\JournalTitle{Structure and infrastructure engineering}}}
  \textbf{\bibinfo{volume}{3}}, \bibinfo{pages}{29--41} (\bibinfo{year}{2007}).

\bibitem{Rawlings2000Tutorial}
\bibinfo{author}{Rawlings, J.~B.}
\newblock \bibinfo{journal}{\bibinfo{title}{Tutorial overview of model
  predictive control}}.
\newblock {\emph{\JournalTitle{IEEE Control Systems}}}
  \textbf{\bibinfo{volume}{20}}, \bibinfo{pages}{38--52}
  (\bibinfo{year}{2000}).

\bibitem{Morari1999Model}
\bibinfo{author}{Morari, M.} \& \bibinfo{author}{Lee, J.~H.}
\newblock \bibinfo{journal}{\bibinfo{title}{Model {{Predictive Control}}:
  {{Past}}, {{Present}} and {{Future}}}}.
\newblock {\emph{\JournalTitle{Computers \& Chemical Engineering}}}
  \textbf{\bibinfo{volume}{23}}, \bibinfo{pages}{667--682}
  (\bibinfo{year}{1999}).

\bibitem{Kouro2009Model}
\bibinfo{author}{Kouro, S.}, \bibinfo{author}{Cort{\'e}s, P.},
  \bibinfo{author}{Vargas, R.}, \bibinfo{author}{Ammann, U.} \&
  \bibinfo{author}{Rodr{\'i}guez, J.}
\newblock \bibinfo{journal}{\bibinfo{title}{Model predictive
  control\textemdash{{A}} simple and powerful method to control power
  converters}}.
\newblock {\emph{\JournalTitle{IEEE Transactions on industrial electronics}}}
  \textbf{\bibinfo{volume}{56}}, \bibinfo{pages}{1826--1838}
  (\bibinfo{year}{2009}).

\bibitem{Qin2003survey}
\bibinfo{author}{Qin, S.~J.} \& \bibinfo{author}{Badgwell, T.~A.}
\newblock \bibinfo{journal}{\bibinfo{title}{A {{Survey}} of {{Industrial Model
  Predictive Control Technology}}}}.
\newblock {\emph{\JournalTitle{Control Engineering Practice}}}
  \textbf{\bibinfo{volume}{11}}, \bibinfo{pages}{733--764}
  (\bibinfo{year}{2003}).

\bibitem{Lin2011Fast}
\bibinfo{author}{Lin, S.}, \bibinfo{author}{De~Schutter, B.},
  \bibinfo{author}{Xi, Y.} \& \bibinfo{author}{Hellendoorn, H.}
\newblock \bibinfo{journal}{\bibinfo{title}{Fast model predictive control for
  urban road networks via {{MILP}}}}.
\newblock {\emph{\JournalTitle{IEEE Transactions on Intelligent Transportation
  Systems}}} \textbf{\bibinfo{volume}{12}}, \bibinfo{pages}{846--856}
  (\bibinfo{year}{2011}).

\bibitem{wonham1964optimal}
\bibinfo{author}{Wonham, W.~M.} \& \bibinfo{author}{Johnson, C.}
\newblock \bibinfo{journal}{\bibinfo{title}{Optimal bang-bang control with
  quadratic performance index}}.
\newblock {\emph{\JournalTitle{Journal of Basic Engineering}}}
  \textbf{\bibinfo{volume}{86}}, \bibinfo{pages}{107--115}
  (\bibinfo{year}{1964}).

\bibitem{Bemporad1999Control}
\bibinfo{author}{Bemporad, A.} \& \bibinfo{author}{Morari, M.}
\newblock \bibinfo{journal}{\bibinfo{title}{Control of {{Systems Integrating
  Logic}}, {{Dynamics}}, and {{Constraints}}}}.
\newblock {\emph{\JournalTitle{Automatica}}} \textbf{\bibinfo{volume}{35}},
  \bibinfo{pages}{407--427} (\bibinfo{year}{1999}).

\bibitem{Suman2006survey}
\bibinfo{author}{Suman, B.} \& \bibinfo{author}{Kumar, P.}
\newblock \bibinfo{journal}{\bibinfo{title}{A {{Survey}} of {{Simulated
  Annealing}} as a {{Tool}} for {{Single}} and {{Multiobjective
  Optimization}}}}.
\newblock {\emph{\JournalTitle{Journal of the operational research society}}}
  \textbf{\bibinfo{volume}{57}}, \bibinfo{pages}{1143--1160}
  (\bibinfo{year}{2006}).

\bibitem{Glover1998Tabu}
\bibinfo{author}{Glover, F.} \& \bibinfo{author}{Laguna, M.}
\newblock \bibinfo{title}{Tabu {{Search}}}.
\newblock In \emph{\bibinfo{booktitle}{Handbook of Combinatorial
  Optimization}}, \bibinfo{pages}{2093--2229} (\bibinfo{publisher}{{Springer}},
  \bibinfo{year}{1998}).

\bibitem{Mitchell1998Handbook}
\bibinfo{author}{Mitchell, M.}
\newblock \bibinfo{journal}{\bibinfo{title}{Handbook of {{Genetic
  Algorithms}}}}.
\newblock {\emph{\JournalTitle{Artif. Intell.}}}
  \textbf{\bibinfo{volume}{100}}, \bibinfo{pages}{325--330}
  (\bibinfo{year}{1998}).

\bibitem{vazirani2013approximation}
\bibinfo{author}{Vazirani, V.~V.}
\newblock \emph{\bibinfo{title}{Approximation {{Algorithms}}}}
  (\bibinfo{publisher}{{Springer Science \& Business Media}},
  \bibinfo{year}{2013}).

\bibitem{OMalley2017Nonnegative}
\bibinfo{author}{O'Malley, D.}, \bibinfo{author}{Vesselinov, V.~V.},
  \bibinfo{author}{Alexandrov, B.~S.} \& \bibinfo{author}{Alexandrov, L.~B.}
\newblock \bibinfo{journal}{\bibinfo{title}{Nonnegative/binary matrix
  factorization with a {{D}}-{{Wave}} quantum annealer}}.
\newblock {\emph{\JournalTitle{arXiv preprint arXiv:1704.01605}}}
  (\bibinfo{year}{2017}).

\bibitem{Ohzeki2018Optimization}
\bibinfo{author}{Ohzeki, M.}, \bibinfo{author}{Okada, S.},
  \bibinfo{author}{Terabe, M.} \& \bibinfo{author}{Taguchi, S.}
\newblock \bibinfo{journal}{\bibinfo{title}{Optimization of neural networks via
  finite-value quantum fluctuations}}.
\newblock {\emph{\JournalTitle{Scientific reports}}}
  \textbf{\bibinfo{volume}{8}}, \bibinfo{pages}{9950} (\bibinfo{year}{2018}).

\bibitem{Tran2016hybrid}
\bibinfo{author}{Tran, T.~T.} \emph{et~al.}
\newblock \bibinfo{title}{A hybrid quantum-classical approach to solving
  scheduling problems}.
\newblock In \emph{\bibinfo{booktitle}{Ninth {{Annual Symposium}} on
  {{Combinatorial Search}}}} (\bibinfo{year}{2016}).

\bibitem{Neven2008Training}
\bibinfo{author}{Neven, H.}, \bibinfo{author}{Denchev, V.~S.},
  \bibinfo{author}{Rose, G.} \& \bibinfo{author}{Macready, W.~G.}
\newblock \bibinfo{journal}{\bibinfo{title}{Training a {{Binary Classifier}}
  with the {{Quantum Adiabatic Algorithm}}}}.
\newblock {\emph{\JournalTitle{arXiv:0811.0416 [quant-ph]}}}
  (\bibinfo{year}{2008}).
\newblock \eprint{0811.0416}.

\bibitem{Neven2009Training}
\bibinfo{author}{Neven, H.}, \bibinfo{author}{Denchev, V.~S.},
  \bibinfo{author}{Rose, G.} \& \bibinfo{author}{Macready, W.~G.}
\newblock \bibinfo{journal}{\bibinfo{title}{Training a {{Large Scale
  Classifier}} with the {{Quantum Adiabatic Algorithm}}}}.
\newblock {\emph{\JournalTitle{arXiv:0912.0779 [quant-ph]}}}
  (\bibinfo{year}{2009}).
\newblock \eprint{0912.0779}.

\bibitem{cairano2007model}
\bibinfo{author}{Cairano, S.~D.}, \bibinfo{author}{Bemporad, A.},
  \bibinfo{author}{Kolmanovsky, I.~V.} \& \bibinfo{author}{Hrovat, D.}
\newblock \bibinfo{journal}{\bibinfo{title}{Model predictive control of
  magnetically actuated mass spring dampers for automotive applications}}.
\newblock {\emph{\JournalTitle{International Journal of Control}}}
  \textbf{\bibinfo{volume}{80}}, \bibinfo{pages}{1701--1716}
  (\bibinfo{year}{2007}).

\bibitem{Mitchell2004Introduction}
\bibinfo{author}{Mitchell, J.~L.}
\newblock \bibinfo{journal}{\bibinfo{title}{Introduction to digital audio
  coding and standards}}.
\newblock {\emph{\JournalTitle{Journal of Electronic Imaging}}}
  \textbf{\bibinfo{volume}{13}}, \bibinfo{pages}{399} (\bibinfo{year}{2004}).

\bibitem{Zwicker2013Psychoacoustics}
\bibinfo{author}{Zwicker, E.} \& \bibinfo{author}{Fastl, H.}
\newblock \bibinfo{journal}{\bibinfo{title}{Psychoacoustics: {{Facts}} and
  models}}.
\newblock {\emph{\JournalTitle{Springer Science \& Business Media}}}
  (\bibinfo{year}{2013}).

\bibitem{Pohlmann:2000:PDA:540747}
\bibinfo{author}{Pohlmann, K.~C.}
\newblock \emph{\bibinfo{title}{Principles of {{Digital Audio}}}}
  (\bibinfo{publisher}{{McGraw-Hill Professional}}, \bibinfo{year}{2000}),
  \bibinfo{edition}{4th} edn.

\bibitem{Wannamaker1992Psychoacoustically}
\bibinfo{author}{Wannamaker, R.}
\newblock \bibinfo{journal}{\bibinfo{title}{Psychoacoustically {{Optimal Noise
  Shaping}}}}.
\newblock {\emph{\JournalTitle{Journal of the Audio Engineering Society}}}
  \textbf{\bibinfo{volume}{40}}, \bibinfo{pages}{611--620}
  (\bibinfo{year}{1992}).

\bibitem{Boothby:2016:FCM:2877060.2877142}
\bibinfo{author}{Boothby, T.}, \bibinfo{author}{King, A.~D.} \&
  \bibinfo{author}{Roy, A.}
\newblock \bibinfo{journal}{\bibinfo{title}{Fast clique minor generation in
  chimera qubit connectivity graphs}}.
\newblock {\emph{\JournalTitle{Quantum Information Processing}}}
  \textbf{\bibinfo{volume}{15}}, \bibinfo{pages}{495--508},
  \doiprefix\url{10.1007/s11128-015-1150-6} (\bibinfo{year}{2016}).

\bibitem{Cai2014practical}
\bibinfo{author}{Cai, J.}, \bibinfo{author}{Macready, W.~G.} \&
  \bibinfo{author}{Roy, A.}
\newblock \bibinfo{journal}{\bibinfo{title}{A practical heuristic for finding
  graph minors}}.
\newblock {\emph{\JournalTitle{arXiv:1406.2741 [quant-ph]}}}
  (\bibinfo{year}{2014}).
\newblock \eprint{1406.2741}.

\bibitem{Hamerlyeaau0823}
\bibinfo{author}{Hamerly, R.} \emph{et~al.}
\newblock \bibinfo{journal}{\bibinfo{title}{Experimental investigation of
  performance differences between coherent {{Ising}} machines and a quantum
  annealer}}.
\newblock {\emph{\JournalTitle{Science Advances}}}
  \textbf{\bibinfo{volume}{5}}, \doiprefix\url{10.1126/sciadv.aau0823}
  (\bibinfo{year}{2019}).
\newblock
  \eprint{https://advances.sciencemag.org/content/5/5/eaau0823.full.pdf}.

\bibitem{W.Johnson2018Future}
\bibinfo{author}{W.~Johnson, M.}
\newblock \bibinfo{title}{Future {{Hardware Directions}} of {{Quantum
  Annealing}}}.
\newblock In \emph{\bibinfo{booktitle}{Qubits {{Europe}} 2018 {{D}}-{{Wave
  Users Conference}}}} (\bibinfo{address}{{Munich}}, \bibinfo{year}{2018}).

\bibitem{zhang2001stability}
\bibinfo{author}{Zhang, W.}, \bibinfo{author}{Branicky, M.~S.} \&
  \bibinfo{author}{Phillips, S.~M.}
\newblock \bibinfo{journal}{\bibinfo{title}{Stability of networked control
  systems}}.
\newblock {\emph{\JournalTitle{IEEE control systems magazine}}}
  \textbf{\bibinfo{volume}{21}}, \bibinfo{pages}{84--99}
  (\bibinfo{year}{2001}).

\end{thebibliography}



\section*{Acknowledgements}
The authors would like to thank Dr.~Akihisa Okada, Dr.~Tadayoshi Matsumori, Dr.~Yuji Ito and Dr.~Kiyosumi Kidono of Toyota Central R\&D Labs. Inc., and Dr. Yoshiki Matsuda of Fixstars Corporation for the useful discussions.




\end{document}